\documentclass[nofootinbib,showpacs,preprintnumbers,amsmath,amssymb,floatfix]
{revtex4-1}
\usepackage{graphicx}

\begin{document}

%\begin{center}

\title{Intrinsic mass scale in QCD factorization}

\vspace*{0.3 cm}

\author{B.I.~Ermolaev}
\affiliation{Ioffe Physico-Technical Institute, 194021
 St.Petersburg, Russia}
\author{S.I.~Troyan}
\affiliation{St.Petersburg Institute of Nuclear Physics, 188300
Gatchina, Russia}

\begin{abstract}
In this paper we argue for existence  of an intrinsic mass scale in QCD factorization
and present a possible origin of it.
Values of this scale are within the Non-Perturbative QCD mass range. It differs from the known factorization scale which
is within the perturbative mass range and dependence on which vanishes in
the factorization convolutions. We show that the intrinsic mass scale plays the key part in
reduction of $K_T$ Factorization to Collinear Factorization: such a reduction can be done
provided that dependence of the non-perturbative inputs in $K_T$ Factorization on invariant
energy has a sharp-peaked form. In this case the intrinsic mass scale is associated with location of the peak(s).
We also present models where the intrinsic scale is generated by such peaks.
\end{abstract}


\maketitle

\section{Introduction}

The concept of QCD factorization suggests that description of hadronic reactions at high energies can be divided
into perturbative and non-perturbative stages. It can be applied to any process involving hadrons.
Among the simplest applications, there are such constructions as the
DIS structure functions, parton distribution in hadrons, etc.
The Optical theorem relates these objects to the Compton scattering amplitudes off hadrons, parton-hadron scattering amplitudes, etc.
 Basically, both
Single-Parton and Multi-Parton Scattering scenarios contribute to
those processes but in the present paper we consider the simplest
and at the same time most popular scenario of the Single-Parton Scattering only. In order to be specific,
we consider the gluon distribution $D$ in hadrons though our conclusions are valid for all hadronic reactions
as long as the Single-Parton scenario is pursued.
The Optical theorem
relates it to the color singlet gluon-hadron scattering amplitude $A$ in the forward kinematics.
Factorization of $A$ is graphically represented in Fig.~\ref{fsfig1} in all available forms of
QCD factorization:

%%%%%%%%%%%%%%%%%%%%%%%%%%%%%%%%%%%%%%%%%%%%
\begin{figure}[h]
  \includegraphics[width=.25\textwidth]{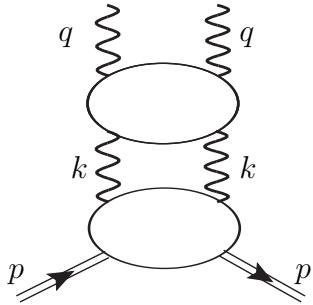}
  \caption{\label{fsfig1} Factorization of amplitudes of hadron-gluon scattering in the
forward kinematics, with intermediate partons being gluons. The
upper blob corresponds to amplitudes of the elastic gluon scattering.}
\end{figure}
%%%%%%%%%%%%%%%%%%%%%%%%%%%%%%%%%%%%%%%%%%%

The upper blob in Fig.~\ref{fsfig1} corresponds to the perturbative amplitude $A^{(pert)}$
which is calculated by perturbative means. In contrast, the lowest blob
is addressed as non-perturbative. It is defined through various models and fits.
It is worth mentioning that Fig.~\ref{fsfig1} includes the graph with non-zero imaginary
part in $s$ ($s = (p + q)^2$) only albeit a similar graph, with $q \leftrightarrows -q$,
also contributes to $A$ but it vanishes when the Optical theorem has been applied.
In both Collinear and $K_T$- Factorization the graph in Fig.~\ref{fsfig1} has a symbolic meaning only and
one cannot apply the Feynman rules to obtain analytic expressions.
In general, the gluon distributions in the hadrons depend on the kinematic variables $s,q^2,p^2$ (see Fig.~\ref{fsfig1})
and on the hadron spin: $D = D (s,q^2, p^2, S_h)$.  For the sake of simplicity, we skip
writing the spin dependence
though our results embrace the case of the polarized distributions either.
In terms of
Collinear Factorization\cite{colfact} $D$ can be represented  as

\begin{equation}\label{colfact}
D (s,q^2, p^2) \approx D_{col} (s,q^2, p^2) = D_{col}^{(pert)} (x/\beta, q^2, \mu^2_{col}) \otimes \phi_{col} (\beta, \mu^2_{col})~,
\end{equation}
in terms of the more general $K_T$-Factorization~\cite{ktfact} its representation is

\begin{equation}\label{ktfact}
D(s,q^2, p^2) \approx D_{KT}(s,q^2, p^2) = D_{KT}^{(pert)} (x/\beta, q^2, k^2_{\perp}, \mu^2_{KT}) \otimes
\Phi_{KT}(\beta, k^2_{\perp}, \mu^2_{KT})
\end{equation}
and in terms of the most general Basic Factorization~\cite{egtfact}

\begin{equation}\label{basfact}
D(s,q^2, p^2) \approx D_{BF}(s,q^2, p^2) = D_{BF}^{(pert)} (x/\beta, q^2, k^2)\otimes \Phi_{BF} (\alpha, k^2, \mu^2_{BF})~.
\end{equation}

In Eqs.~(\ref{colfact},\ref{ktfact},\ref{basfact}) the notations $D_{col}^{(pert)}, D_{KT}^{(pert)}$ and $D_{BF}^{(pert)}$ stand
for perturbative contributions while $\phi_{col}$ and $\Phi_{KT}$  are integrated and unintegrated
gluon distributions respectively. They are supposed to accommodate both perturbative and non-perturbative
contributions. In
 contrast, the totally unintegrated gluon distribution $\Phi_{BF}$ contains non-perturbative contributions only.
 The symbol $\otimes$ refers to the different integrations, depending on the form of factorization.
 In Collinear Factorization it means the integration over the longitudinal
fraction $\beta$ of momentum $k$.
In $K_T$-Factorization the symbol $\otimes$ means the two-dimensional integration:  over both $\beta$ and transverse momentum
$k_{\perp}$.
In Basic Factorization it denotes the three-dimensional integration: besides integrations over $\beta$ and $k_{\perp}$,
it involves integration over the second longitudinal variable, $\alpha$ dependence on which is left unaccounted in Collinear and $K_T$
Factorization.
%The integration regions in Eqs.~(\ref{colfact},\ref{ktfact},\ref{basfact}) can be found in Appendix A.
The variables $\alpha, \beta, k_{\perp}$ are related to the Sudakov parametrization\cite{sud} of momentum $k$:

\begin{equation}\label{sud}
k = - \alpha q' + \beta p' + k_{\perp}~,
\end{equation}
where the light-cone momenta $p',q'$ are made of the external momenta $p$ and $q$:
%\begin{equation}\label{pq}
%p' = p - x_p q,~q' = q - x_q p,~ x_p \approx |p^2|/w,~x_q \approx |q^2|/w,~w = 2p'q' \approx 2pq.
%\end{equation}
\begin{equation}\label{pq}
p = p' + x_p q',~q = q' - x_q p',~ x_p = p^2/w,~x_q = q^2/w,~w = 2p'q' \approx 2pq~.
\end{equation}

The parameters $\mu_{col}$ and $\mu_{KT}$ are called the factorization scale in Collinear and $K_T$-
Factorizations respectively. Let us note that $\mu_{col}$ is the only mass scale
for the integrated parton distributions $\phi_{col}$ while situation in $K_T$- Factorization is more involved.
The distributions $\phi_{col}$ and $\Phi_{KT}$ depend on the factorization scale implicitly, through
phenomenological numerical factors. Dependence of  $\phi_{col}$ and $\Phi_{KT}$ on the factorization scale
is exactly compensated by the inverse dependence of the perturbative
%the $\mu_{col}$ and $\mu_{KT}$ -dependence of the perturbative
contributions in both Eqs~(\ref{colfact}) and (\ref{ktfact}),
%respectively.
which can be interpreted as if
Eqs~(\ref{colfact},\ref{ktfact}) were free of any mass scale at all.
In contrast,
dependence of $\Phi_{BF}$ on $\mu_{BF}$ is explicit, it does not vanish in the convolution (\ref{basfact}), which
leads to implicit dependence of $D_{BF}$ on $\mu_{BF}$. The scale $\mu_{BF}$ is an intrinsic mass scale in Basic Factorization.

In the present paper we prove that dependence on the intrinsic mass scale
exists in  both Collinear and
$K_T$- Factorizations. It does not vanish in factorization convolutions. We argue that typical
values of this scale are $\sim \Lambda_{QCD}$, i.e. they are in the domain of Non-Perturbative QCD.
 Our paper is organized as follows: In Sect.~II we consider the problem of the intrinsic mass scale
 in the conventional forms of factorization while in Sect.~III we study the same problem in the
 framework of the more general form, Basic Factorization. We show in Sects.~II,III that
 no source for the intrinsic mass scale
 can be found when all those forms of QCD factorization are regarded as unrelated to each other.
 In Sect.~IV we explain how to
 reduce Basic Factorization to $K_T$ Factorization and then in Sect.~IV we reduce $K_T$ Factorization to Collinear Factorization.
By doing so we find out a possible source of the intrinsic mass scale and study it in general.
 In Sect.V we suggest a simple model involving the intrinsic mass scale. Sect.~VI is for our concluding remarks.

\section{Conventional treatment of the mass/factorization scale in QCD factorization}

Treatments of the mass scale in Collinear and $K_T$- Factorizations are much alike. On the other
hand,
Collinear Factorization is the simplest form of QCD factorization, so in the first place
we consider handling the factorization scale $\mu_{col}$  in Eq.~(\ref{colfact}).
Collinear Factorization was designed so that
the perturbative contribution $D_{col}^{(pert)}$ could be
calculated with the DGLAP equations\cite{dglap}, with the input $\phi_{col}(x_0, \mu^2_{col})$ being
defined at the mass scale $\mu^2_{col}$ and at $x_0 \sim 1$. Values of $\mu_{col}$ are conventionally high enough:
$\mu_{col} \sim$ few GeV. It keeps the perturbative contribution $D_{col}^{(pert)}$
fairly within the domain of Perturbative QCD. At the same time, the integrated parton
density $\phi_{col}(x_0,\mu^2_{col})$, being defined at such high scale, cannot be free of
perturbative contributions, so in addition to non-perturbative contributions it accommodates also
perturbative contributions. Because of that $\phi_{col}(x_0,\mu^2_{col})$ can be regarded as one obtained
with the same perturbative evolution of the input defined at a lower scale $\mu$.
Relation between the inputs defined at the scales $\mu^2_{col}$ and $\mu^2$ can be written symbolically as follows:

\begin{equation}\label{phimu}
\phi_{col}(x_0, \mu^2_{col}) = E_{DGLAP} (\mu^2_{col},\mu^2) \otimes \phi_{col}(x_0, \mu^2),
\end{equation}
where $\phi_{col}(x_0, \mu^2)$ is the non-perturbative input and the integral operator $E_{DGLAP} (\mu^2_{col}, \mu^2) $,
with $\mu^2_{col}$ and $\mu^2$ being the upper and lower limits of the integration respectively,
 is made of the splitting functions. After applying the Mellin transform to Eq.~(\ref{phimu}),
 the operator $E_{DGLAP}$ is expressed through the anomalous dimensions.
Substituting Eq.~(\ref{phimu}) in Eq.~(\ref{colfact}), we can express $D_{col}$ through  $\phi_{col}(x_0, \mu^2)$:

\begin{equation}\label{colfactmu}
D_{col} (s,q^2, p^2) = D_{col}^{(pert)} (x/\beta, q^2, \mu^2_{col}) \otimes
E_{DGLAP} (\mu^2_{col},\mu^2) \otimes \phi_{col}(x_0, \mu^2).
\end{equation}

The convolution $D_{col}^{(pert)} (x/\beta, q^2, \mu^2_{col}) \otimes
E_{DGLAP} (\mu^2_{col},\mu^2)$ represents the perturbative contribution defined at the same $x_0$ and the new scale $\mu$:

\begin{equation}\label{dpertmu}
D_{col}^{(pert)} (x/\beta, q^2, \mu^2_{col}) \otimes
E_{DGLAP} (\mu^2_{col},\mu^2) = D_{col}^{(pert)} (x/\beta, q^2, \mu^2)
\end{equation}

and therefore
we arrive at the expression for $D_{col}$, where the input is defined at the lesser scale $\mu$.
If $\mu$ is also in the Perturbative domain, we can apply Eq.~(\ref{phimu}) to
$\phi_{col}(x_0, \mu^2)$, and doing so we eventually arrive at $D_{col}$, where the input is defined at the minimal scale $\mu^2_0$:

\begin{equation}\label{dmuzero}
D_{col} (s,q^2, p^2) = D_{col}^{(pert)} (x/\beta, q^2, \mu^2_0) \otimes \phi_{col}(x_0, \mu^2_0).
\end{equation}

The scale $\mu_0$ should adjoin
the non-perturbative domain of QCD so that the perturbative evolution could not start from a lesser
scale.
According to the concept of QCD factorization, convolutions of perturbative and non-perturbative contributions do not depend on the factorization point $\mu_{col}$. For instance, $D_{col}$ in Eq.~(\ref{colfact}) should not depend on $\mu_{col}$.
Eqs.~(\ref{phimu},\ref{dpertmu}) make it obvious. On the other hand, Eq.~(\ref{dmuzero}) reads that
$D_{col}$ acquires dependence on $\mu_0$ which is the minimally possible starting point of
perturbative evolution. Such dependence is not present explicitly in the DGLAP fits (see e.g. Ref.~\cite{fits})
available in the literature though it was implied implicitly. We call $\mu_0$ the intrinsic mass scale.

DGLAP was constructed to operate in the region of large $x$, where the longitudinal and transverse sub-spaces
are approximately factorized, i.e. evolutions in $k$ and $k_{perp}$ are independent.
When $x \ll 1$, such factorization breaks and DGLAP should not be applied. The point is
that DGLAP cannot account for contributions $ \sim \ln^n x$ which becomes important at small $x$
and should be resummed to all orders in $\alpha_s$. Besides, introducing the scale $\mu_0$
in the small-$x$ region is important because it acts as a cut-off for the
infrared-divergent double-logarithmic contributions.
The impact of them replaces the operator $E_{DGLAP}(x_0, \mu^2_2,\mu^2_1)$ by the improved operator $E(x_2,x_1, \mu^2_2,\mu^2_1)$, which changes the Eq.~(\ref{phimu}) and allows one to
combine the evolutions in both $x$ and $k_T$:

\begin{equation}\label{phimux}
\phi_{col}(x, \mu^2_{col}) = E (x,x_0, \mu^2_{col},\mu^2) \otimes \phi_{col}(x_0, \mu^2)~.
\end{equation}

However, this circumstance does not affect the final result and Eq.~(\ref{dmuzero}) is valid in
the small-$x$ region as well.
We remind that the generalization of DGLAP to include the regions of the both large and small $x$
was suggested in
Refs.~\cite{egtns}, then it was extended on the region of small $Q^2$ in Ref.~\cite{egtsmallq},
see also the overviews~\cite{egtg1sum}.
The treatment of the QCD mass scale in $K_T$-Factorization is absolutely the same like in Collinear Factorization.
So, we have demonstrated that both  $K_T$- and Collinear Factorizations imply existence of some primary
mass scale $\mu_0$ whose value is close to the Non-Perturbative DCD domain.
Because of that we call $\mu_0$ the intrinsic mass scale.
On the other hand, all available phenomenological fits do not include $\mu_0$ explicitly and do not
give any hint on how it could appear. Below we show that the intrinsic mass scale $\mu_0$ does originate quite
naturally and in a model-independent way in Basic Factorization.

\section{Reminding basic facts about Basic Factorization}

\begin{figure}[h]
  \includegraphics[width=.15\textwidth]{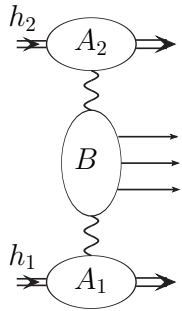}
  \caption{\label{fsfig2} Amplitude for the Single-Parton Scattering of hadrons $h_1$ and $h_2$ , with active partons
being gluons. Blobs $A_{1,2}$ denote emission of the active gluons.
Interaction of those gluons is depicted by blob $B$, where the outgoing
arrows denote the produced partons. The outgoing double arrows on blobs $A_{1,2}$ stand for the final state
spectators.}
\end{figure}

Basic Factorization suggested in Ref.~\cite{egtfact} is the most recent and the most general of the known forms of QCD factorization.
It has been considered in the Single-Parton Scattering approximation only but it can easily be extended to the
Multi-Parton Scattering. As this form of factorization is much less known
than Collinear and $K_T$-Factorizations, we briefly remind below its essence and then proceed to scrutinizing the problem of intrinsic mass.
Its derivation is simple and based on the following observation: each of the
colliding hadrons emits one of several active partons (quarks or gluons) which interact and produce new partons. This interaction is described by Perturbative QCD.
The most popular approximation is that every colliding hadron emits only one active parton. This scenario is called
Single-Parton Scattering and we focus on it. The scattering amplitude of those process is
depicted in Fig.~\ref{fsfig2} for the case when the active partons are gluons.

%%%%%%%%%%%%%%%%%%%%%%%%%%%%%%%%%%%%%%%%%%%%
%%%%%%%%%%%%%%%%%%%%%%%%%%%%%%%%%%%%%%%%%%%

Applying this reasoning to the gluon-proton collision and convoluting its scattering amplitude with the mirror
graph, we arrive at the graph with the two-parton state in $t$-channel. An example of such graphs, with
the active partons being gluons, is given in Fig.1, where the $s$-channel cut is implied.
We have already reminded that factorization convolutions are illustrated in the literature by the same pictures
 regardless of the form of QCD factorization. However, one cannot apply the standard Feynman rules to those pictures
for obtaining analytic expressions in both
Collinear and $K_T$- Factorizations. In contrast, such pictures become the Feynman graphs in Basic Factorization and one
can obtain proper analytic expressions from them, using the standard Feynman rules. Applying the Feynman rules to the
graph in Fig.~\ref{fsfig1}, it is easy to arrive at the corresponding analytic expression for gluon-hadron scattering amplitude $A$ in the forward scattering kinematics.
The expressions for $A$ in the case of both polarized and unpolarized hadrons are obtained in Ref.~\cite{etgluon}.
For instance, in the case of unpolarized hadrons the gluon-hadron scattering amplitude $A$ is

\begin{equation}\label{agen}
A = - \imath \int \frac{d \beta}{\beta} d k^2_{\perp} d \alpha A^{(pert)} (s_2,q^2,k^2)\frac{w}{k^2 k^2} M(s_1,k^2),
\end{equation}

where

\begin{equation}\label{s12}
s_1 = (p-k)^2 \approx w \alpha + k^2 + p^2~,\qquad s_2 = (q + k)^2 \approx w \beta + q^2 + k^2~,\qquad  k^2 = - w \alpha\beta - k^2_{\perp}~.
\end{equation}

The factors $k^2$ in the denominator correspond to propagators of the active gluons.
The notation $A^{(pert)}$
stands for the perturbative gluon-gluon scattering amplitude (the upper blob in Fig.~\ref{fsfig1} and $M$ is the
non-perturbative input. It corresponds to the lowest blob in Fig.~\ref{fsfig1}. The renormalization
makes the perturbative amplitude
$A^{(pert)}$ be free of ultraviolet (UV) divergences whereas infrared (IR) divergences are regulated by
virtualities of the external momenta $q$ and $k$ providing that $q^2 \neq 0, k^2 \neq 0$.
However, even with regulated $A^{(pert)}$, the integrand
of Eq.~(\ref{agen}) can be divergent at $k^2 = 0$ because the integration
runs over the whole phase space. Similarly, integration over $\alpha$ can yield divergent result at large
$|\alpha|$. In Ref.~\cite{egtfact,etgluon} we proved that those singularities are killed when the input $M$
satisfies the following requirements:

\begin{equation}\label{mir}
M \sim \left(k^2\right)^{1 + \eta},
\end{equation}
with $\eta >0$, and
\begin{equation}\label{muv}
M \sim \alpha^{-\kappa},
\end{equation}

with $\kappa > 0$, at  $|\alpha|$. These restrictions are valid for the cases of both polarized and unpolarized hadrons.
 They can be regarded as criteria of validity for all models of the inputs $M$. Applying the Optical theorem to
 Eq.~(\ref{agen}), we obtain the gluon parton distribution $D_{BF}$ in Basic Factorization (cf. Eq.~(\ref{basfact})).
 In Ref.~\cite{egtfact,etgluon} we showed that Basic Factorization can be reduced to $K_T$- Factorization.
 To this end, let us notice that when integration over $\alpha$ in Eq.~(\ref{agen}) has been performed,
 the resulting convolution contains integrations over $\beta$ and $k_{\perp}$ and therefore it looks as a
 convolution in $K_T$- Factorization. However, the problem is that in order to reduce Eq.~(\ref{agen})
 to the one of $K_T$- Factorization, one should take $A^{(pert)}$ out of integral over $\alpha$,
 which cannot be done straightforwardly because $A^{(pert)} (s_2,q^2,k^2)$
 depends on $\alpha$ through $k^2 = - w \alpha \beta -k^2_{\perp}$.
 It is possible only approximately if the integration runs over subregion

 \begin{equation}\label{ak}
 |\alpha| \ll k^2_{\perp}/(w \beta)~.
 \end{equation}

Then $k^2 \approx - k^2_{\perp}$, so the only $\alpha$- dependent factor in the integrand of Eq.~(\ref{agen})
is $M$. Integrating it over the region of Eq.~(\ref{ak}), we arrive at the gluon-hadron scattering amplitude $A_{KT}$ in $K_T$- Factorization:

\begin{equation}\label{akt}
A_{KT} = - \imath \int \frac{d \beta}{\beta} d k^2_{\perp} A^{(pert)} (s_2,q^2,k^2_{\perp})\frac{1}{k^2_{\perp}k^2_{\perp}}
\widetilde{M}_{KT}(\beta, k^2_{\perp}),
\end{equation}
where the input $\widetilde{M}_{KT}$ is

\begin{equation}\label{mkt}
\widetilde{M}_{KT} (\zeta,  k^2_{\perp}) = w \int_{- \zeta}^{\zeta} d \alpha M (s_1,k^2_{\perp}) .
\end{equation}
The notation $\zeta$ in Eq.~(\ref{mkt}) stands for the invariant energy of $\widetilde{M}_{KT}$:

\begin{equation}\label{zeta}
\zeta = \xi k^2_{\perp}/\beta,
\end{equation}
with $0 < \xi \ll 1$ while invariant sub-energies $s_{1,2}$ are defined in Eq.~(\ref{s12}). $\widetilde{M}_{KT}$ is the non-perturbative input to
the gluon-hadron scattering amplitude in $K_T$ Factorization. According to
Eq.~(\ref{mir}), $\widetilde{M}_{KT} \sim (k^2_{\perp})^{1 + \eta}$ at small $k^2_{\perp}$.
Let us define $M_{KT}$ via $\widetilde{M}_{KT}$ as

\begin{equation}\label{phikt}
\widetilde{M}_{KT} = k^2_{\perp} M_{KT} (\zeta,k^2_{\perp} )~.
\end{equation}
Using Eq.~(\ref{phikt}) and noticing that $A^{(pert)} (s_2,q^2,k^2_{\perp})$ is dimensionless,
we can rewrite Eq.~(\ref{akt}) in the following form:

\begin{equation}\label{akt1}
A_{KT} =  \int \frac{d \beta}{\beta} \frac{d k^2_{\perp}}{k^2_{\perp}} A^{(pert)}
(x/\beta,  q^2/\zeta) M_{KT}(\zeta, k^2_{\perp}).
\end{equation}

%\begin{equation}\label{akt1}
%A_{KT} =  \int \frac{d \beta}{\beta} \frac{d k^2_{\perp}}{k^2_{\perp}} A^{(pert)}
%(x/\beta, q^2/k^2_{\perp}) M_{KT}(\zeta, k^2_{\perp})~.
%\end{equation}

Applying the Optical theorem to Eq.~(\ref{akt1}), we arrive at the
gluon distribution in $K_T$ Factorization:

\begin{equation}\label{dkt}
D_{KT} = \int \frac{d \beta}{\beta} \frac{d k^2_{\perp}}{k^2_{\perp}} D^{(pert)} (x/\beta, q^2/\zeta) \Phi_{KT}(\zeta, k^2_{\perp}),
\end{equation}

%\begin{equation}\label{dkt}
%D_{KT} = \int \frac{d \beta}{\beta} \frac{d k^2_{\perp}}{k^2_{\perp}} D^{(pert)} (x/\beta, q^2/k^2_{\perp}) \Phi_{KT}(\zeta, k^2_{\perp})~,
%\end{equation}

where $D^{(pert)}$ is the perturbative contribution and $\Phi_{KT}$ is the non-perturbative input.
This expression coincides with the conventional expression in Eq.~(\ref{ktfact}).
We conclude that introducing Basic Factorization and reducing it to $K_T$
Factorizations do not involve any mass scale, so the problem of
 origin of the intrinsic mass scale still remains unclear.
However, existence of the intrinsic mass scale plays the key role, when  $K_T$-Factorization is reduced to Collinear one.

\section{Reducing $\textbf{K}_\textbf{T}$- Factorization to Collinear Factorization}

\begin{figure}[h]
\centerline{\includegraphics[width=.4\textwidth]{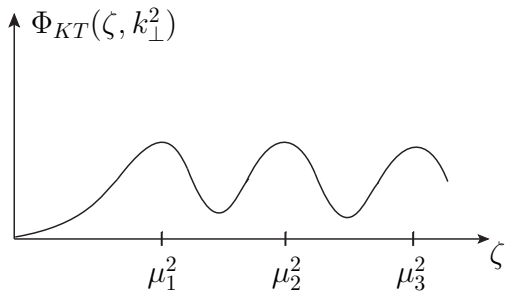}}
\caption{\label{fsfig3} The sharp-peaked form
of dependence of $\Phi_{KT}$ on $\zeta$~.
The number of the maximums is unlimited.
Their widths and heights can be different.}
\end{figure}

Convolutions in Collinear Factorization involve one-dimension integration over $\beta$. Nevertheless, one cannot
arrive at an expression for the gluon-hadron scattering amplitude $A_{col}$ in Collinear Factorization with the
straightforward integration of Eq.~(\ref{akt1}) over $k_{\perp}$ because such integration inevitably
involves integrating $A^{(pert)}$.
% because $A^{(pert)}$ depends on $k_{\perp}$explicitly.
 The only approximate way to integrate out the $k_{\perp}$- dependence without involving $A^{(pert)}$
 is to presume that the $\zeta$-dependence of
$\Phi_{KT}(\zeta, k^2_{\perp})$ has a sharp-peaked form. Generally speaking, the number of the peaks can be unlimited, their widths and heights
can be different. An example of such $\zeta$ -dependence is shown in Fig.~\ref{fsfig3}.

%%%%%%%%%%%%%%%%%%
%%%%%%%%%%%%%%%%%%

In what follows we will address the peaks as
resonances. The suggestion of having one or several resonances in $\Phi_{KT}(\zeta, k^2_{\perp})$
is well grounded. Indeed, after the active quark(s) has been emitted off
the hadron, the remaining ensemble of spectators becomes unstable, so it can likely be described in terms of resonances.
As $\Phi_{KT}$ is altogether non-perturbative, all $\mu_r$ should be within the non-perturbative domain
($\mu_r \sim \Lambda_{QCD}$).
Another alternative is to interpret the peak
scenario in Fig.~\ref{fsfig3} is to represent $\Phi_{KT}$ as a periodic non-perturbative structure.
For instance, it seems possible to express $\Phi_{KT}$ in terms of soliton contributions.
As soon as we accept the peaked form of $\Phi_{KT}(\zeta, k^2_{\perp})$, we can integrate
over $\zeta$ the input $\Phi_{KT}$ only, arriving at the explicit expression for
the gluon-hadron scattering amplitude $A_{(col)}$ in Collinear Factorization. It consists of the resonance and background contributions:

\begin{equation}\label{acolrb}
A_{col} = A_R + A_B~.
\end{equation}
The resonance contribution $A_R$ is
\begin{equation}\label{ar}
A_{R} \approx\sum_n\int\frac{d\beta}{\beta}A^{(pert)}(x/\beta,\mu^2_n)\, \varphi_{R}^{(n)}(\beta, \mu^2_n)~,
\end{equation}
with $\mu^2_n$ being the location of the $n_{th}$ maximum and

\begin{equation}\label{phir}
\varphi_{R}(\beta, \mu^2_n)
\approx\int_{\Omega_n}d\zeta\,\Phi_{KT}(\zeta, k^2_{\perp})~,
\end{equation}
where the integration region $\Omega_n$ is located around the maximum of the $n_{th}$ peak.
Apparently, the regions $\Omega_n$ in Eq.~(\ref{phir}) are not well-defined and they are only a part of the total
integration region. So besides the resonance contribution $A_R$, there are additional non-factorized contributions
which we interpret as background contribution $A_B$ to the amplitude $A$.  Generally speaking, impact of them strongly depend
on the specific expressions for $\Phi_{KT}$.

We have shown that reduction of $K_T$ Factorization down to Collinear Factorization requires existence
of at least one mass scale located in the non-perturbative domain. We call such scales $\mu_n$ the intrinsic mass scales.
These scales have a physical meaning: $\Phi_{KT}$ has the maximums at $\zeta = \mu^2_n$.
Below we present a simple model for $\Psi_{KT}$.

\section{Minimal Resonance Model for $\textbf{K}_\textbf{T}$ Factorization}

Models where $\Phi_{KT}$ depends on $\zeta$ in a way shown in Fig.~\ref{fsfig3} were not discussed in literature until we presented the Resonance Model in Ref.~\cite{egtquark}. This model is based on the following observation:
After the active quark has been emitted off the hadron, the remaining set of spectators pick up a color and thereby it
becomes unstable. This observation guides us to model $\Phi_{KT}$ through interference of several resonances.
We represent $\Phi_{KT}$ in the following way:

\begin{equation}\label{phiktry}
\Phi_{KT} (\zeta, k^2_{\perp}) = R_{KT}(k^2_{\perp})\, Y_{KT} (\zeta)~.
\end{equation}

The only rigorous knowledge on $R_{KT}$ is that $R_{KT} \sim (k^2_{\perp})^{\eta}$ at small $k^2_{\perp}$, which follows
from Eq.~(\ref{mir}). On the other hand, $R_{KT}$ should decrease at large $k^2_{\perp}$, which is often
achieved through exponential factors, see e.g. Refs.~\cite{golec,jung,pumplin}. Being motivated
by these models, we
choose $R_{KT}$ as follows:

\begin{equation}\label{rkt}
R_{KT} = N \left(k^2_{\perp}\right)^{\eta} e^{- k^2_{\perp}/k^2_0}~,
\end{equation}
with $N$ being a constant and $k^2_0$ being a parameter.
When $R_{KT}$ decreases so fast, $Y_{KT}$ can be chosen as one of several factors of the Breit-Wigner type:

\begin{equation}\label{ykt}
Y_{KT} = \sum_n \frac{\Gamma_n}{(\zeta - \mu^2_n)^2 + \Gamma^2_n}~,
\end{equation}
where the number of such factors is unlimited. In order to get sharp peaks in Eq.~(\ref{ykt}) we
presume that $\Gamma_n \ll \mu^2_n$, which is standard for the Breit-Wigner expressions.
We call Minimal Resonance Model (MRM) the case when only one resonance is
involved and therefore in this model

\begin{equation}\label{phimrm}
\Phi_{KT} (\zeta, k^2_{\perp}) \approx  \Phi_{MRM} (\zeta, k^2_{\perp})
= R_{KT} (k^2_{\perp})\frac{\Gamma_1}{(\zeta - \mu^2_1)^2 + \Gamma^2_1}~.
\end{equation}

Replacing $\Phi_{KT}$ in Eq.~(\ref{dkt}) by $\Phi_{MRM}$ and integrating over $\zeta$, we obtain the gluon distribution
in Collinear Factorization in the form of Eq.~(\ref{acolrb})

\begin{equation}\label{arb1}
A_{col} = A_R^{(1)} + A_B^{(1)}
\end{equation}

 but with different resonance and background
contributions. The resonance contribution $A_R^{(1)}$ looks very simple:

\begin{equation}\label{ar1}
A_{R}^{(1)} = \int_x^1\frac{d\beta}{\beta}A^{(pert)}(x/\beta, q^2/\mu^2_1)\, \varphi_{R}^{(1)}(\beta, \mu^2_1)~,
\end{equation}
with

\begin{equation}\label{phir1}
\varphi_{R}^{(1)} = \pi N  \left(\frac{\beta \mu^2_1}{k^2_0}\right)^{\eta} e^{- \beta \mu^2_1/k^2_0}
\end{equation}
while the background contribution $A_B^{(1)}$ is given by a more involved expression:
\begin{equation}\label{ab1}
A_B^{(1)} = \sum_{n=1}^{\infty}\frac{N}{n!}   \left(\frac{\Gamma_1}{\mu^2_1}\right)^n
C_n(q^2/\Gamma, \mu^2_1/\Gamma)
\int_x^1 \frac{d \beta}{\beta}
\left[\frac{\partial^n}{\partial y^n } \left(\frac{A^{(pert)}(x/\beta, q^2/(y \mu^2_1))}{y
}\right)\right]_{y = 1}
\left(\frac{\beta \mu^2_1}{k^2_0}\right)^{\eta} e^{- \beta \mu^2_1/k^2_0}~,
\end{equation}
with

\begin{equation}\label{cn}
C_n = \int_{- \mu^2_1/\Gamma}^{q^2/\Gamma} dt \frac{t^n}{1 + t^2}~.
\end{equation}

Let us discuss these results. Eqs.~(\ref{phir1},\ref{ab1}) contain the parameters $\mu^2_1$ and $k^2_0$. Their values are widely different:
$\mu^2_1$ is in the non-perturbative domain, $\mu_1 \sim \Lambda_{QCD}$ whereas $k^2_0$ of Refs.~\cite{golec, jung,pumplin}
is much greater. Therefore, the exponential factor in Eqs.~(\ref{phir1},\ref{ab1}) can be either neglected or approximated by few terms of the
power expansion. It means that the $\beta$-dependence in Eq.~(\ref{phir1}) (and similarly in Eq.~(\ref{ab1})) is approximately $\sim \beta^\eta (1 - \beta (\mu^2_1/k^2_0))$,
where $\eta$ is positive.  Then, the factors $C_n$ in Eq.~(\ref{ab1}) are $\sim \ln^n (q^2/ \Gamma_1)$ and $ \ln^n (\mu^2_1/\Gamma_1)$.
Despite they grow with increase of $n$, the power factors $(\Gamma/ \mu^2_1)^n$ decrease at the same time much faster, so
the series of Eq.~(\ref{ab1}) can be approximated by only few first terms.

\section{Conclusion}

In present paper we have scrutinized the problem of a possible origin of the intrinsic mass scale in  QCD factorization.
To begin with, we demonstrated that all available forms of QCD factorization implicitly need a mass scale, but nothing definite can be deduced from the analysis of factorization when they are studied independently of each other. Investigating relations between different forms of factorization, we found out that reduction of Basic Factorization to $K_T$ Factorization is insensitive to the problem of the intrinsic mass scale. This reduction is done with purely mathematical means: restriction of the phase space. In contrast, reduction of $K_T$ Factorization down to Collinear Factorization is based on physical assumptions. Namely, the non-perturbative inputs  $\Phi_{KT}$ should depend on the invariant energy in a specific way. As the first alternative, there can be a sharp-peeked dependence, where the number of the peaks is unlimited whereas their heights and widths can be different.
In this case, the intrinsic mass scale(s) are associated with location of the peaks. The second option is to have a periodic-function dependence.
%Here $\Phi_{KT}$, being non-perturbative structure,  can probably be represented  through a set of solitons
%but we leave this case uninspected, focusing on the peaked dependence only.
We model the peaked structure of $\Phi_{KT}$  by the series of resonances and, as the simplest case, by the single resonance.
Using the Breit-Wigner expressions, we arrive at representation of the non-perturbative inputs in Collinear
Factorization as the sum of the resonance  contributions and background.

\section{Acknowledgement}

We are grateful to S.~Catani and W.~Schafer for useful communications.

\end{document}